%% file: main.tex
\documentclass[sigconf,authorversion]{acmart}

\usepackage{enumitem}
\usepackage{subcaption}
\usepackage{graphicx}
\usepackage{changepage}
\usepackage{gensymb}
\usepackage{xcolor}

\definecolor{MapKey_PinBlue}{HTML}{009BBA}
\definecolor{MapKey_Pink}{HTML}{FF83F7}
\definecolor{MapKey_Green}{HTML}{008261}
\definecolor{MapKey_Blue}{HTML}{000FCE}

\AtBeginDocument{%
  \providecommand\BibTeX{{%
    \normalfont B\kern-0.5em{\scshape i\kern-0.25em b}\kern-0.8em\TeX}}}

\copyrightyear{2024}
\acmYear{2024}
\setcopyright{acmlicensed}\acmConference[CHI '24]{Proceedings of the CHI Conference on Human Factors in Computing Systems}{May 11--16, 2024}{Honolulu, HI, USA}
\acmBooktitle{Proceedings of the CHI Conference on Human Factors in Computing Systems (CHI '24), May 11--16, 2024, Honolulu, HI, USA}
\acmDOI{10.1145/3613904.3642615}
\acmISBN{979-8-4007-0330-0/24/05}

\begin{document}

\title[Surveyor: Facilitating Discovery Within Video Games for BLV Players]{Surveyor: Facilitating Discovery Within Video Games for Blind and Low Vision Players}


\author{Vishnu Nair}
\email{nair@cs.columbia.edu}
\affiliation{%
 \institution{Columbia University}
 \city{New York}
 \state{New York}
 \country{USA}
}

\author{Hanxiu 'Hazel' Zhu}
\email{hz2653@columbia.edu}
\affiliation{%
 \institution{Columbia University}
 \city{New York}
 \state{New York}
 \country{USA}
}

\author{Peize Song}
\email{ps3366@columbia.edu}
\affiliation{%
 \institution{Columbia University}
 \city{New York}
 \state{New York}
 \country{USA}
}

\author{Jizhong Wang}
\email{jw4153@columbia.edu}
\affiliation{%
 \institution{Columbia University}
 \city{New York}
 \state{New York}
 \country{USA}
}

\author{Brian A. Smith}
\email{brian@cs.columbia.edu}
\affiliation{%
 \institution{Columbia University}
 \city{New York}
 \state{New York}
 \country{USA}
}

\renewcommand{\shortauthors}{Nair, Zhu, Song, Wang, and Smith}

\begin{abstract}
  Video games are increasingly accessible to blind and low vision (BLV) players, yet many aspects remain inaccessible. One aspect is the joy players feel when they explore environments and make new discoveries, which is integral to many games. Sighted players experience discovery by surveying environments and identifying unexplored areas. Current accessibility tools, however, guide BLV players directly to items and places, robbing them of that experience. Thus, a crucial challenge is to develop navigation assistance tools that also foster exploration and discovery. To address this challenge, we propose the concept of exploration assistance in games and design Surveyor, an in-game exploration assistance tool that enhances discovery by tracking where BLV players look and highlighting unexplored areas. We designed Surveyor using insights from a formative study and compared Surveyor's effectiveness to approaches found in existing accessible games. Our findings reveal implications for facilitating richer play experiences for BLV users within games.
\end{abstract}



\begin{teaserfigure}
  \includegraphics[width=0.85\textwidth]{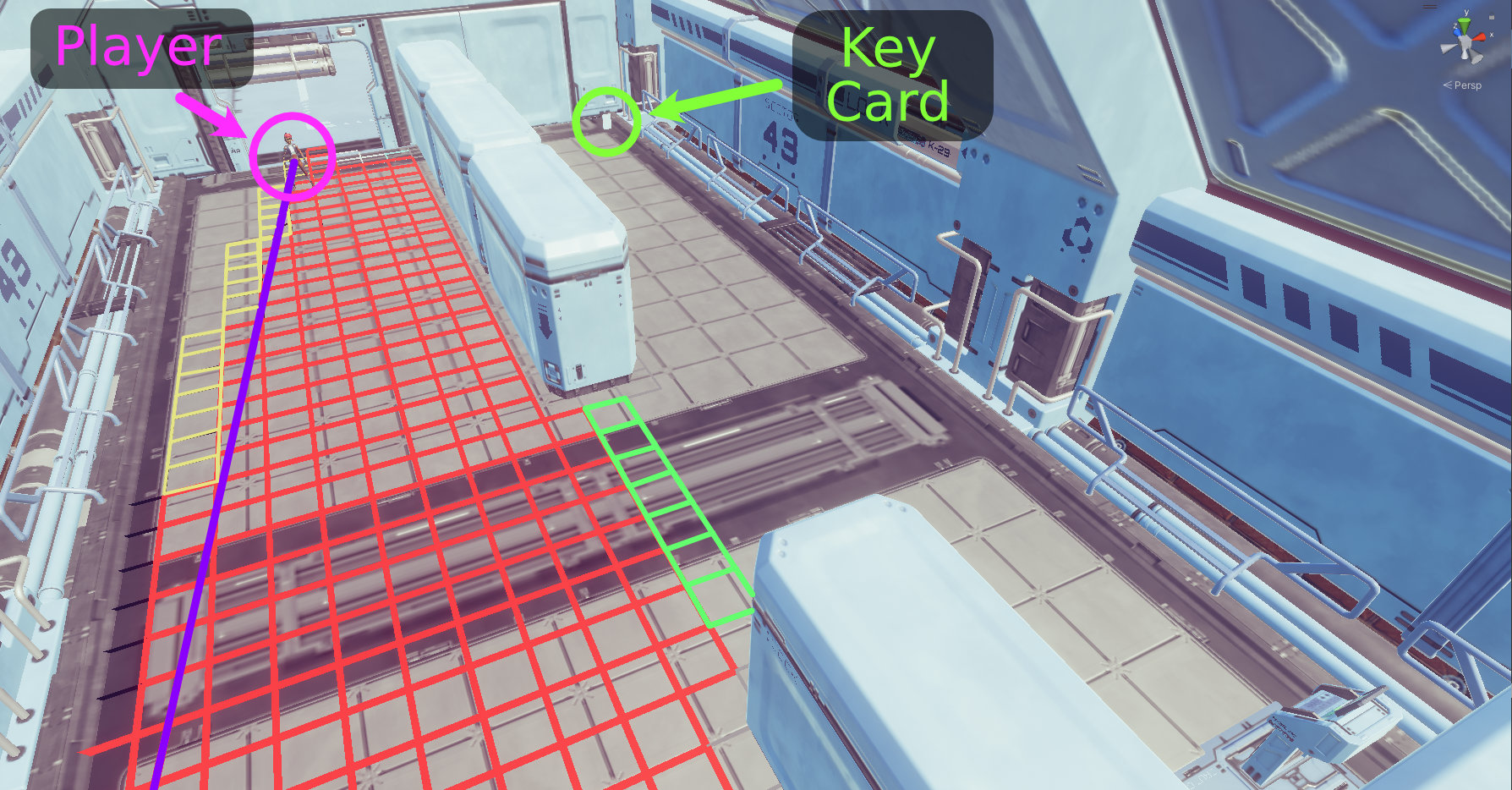}
  \centering
  \caption{View of a player using Surveyor for exploration in our game testbed. Surveyor is an exploration assistance tool designed to foster a sense of discovery in BLV players. Here, the player is using the right stick to survey the area via line-of-sight (purple ray), with red cells marking surveyed areas. Yellow and green cell blocks ("patches") denote boundaries between seen and unseen regions. When the player queries Surveyor's menus, Surveyor highlights each patch's distance and direction, and can guide the player to these areas for further exploration. Here, the key card lies behind large crates, initially unseen. However, if the player makes their way to the green patch of cells and then uses the right stick to survey, they will discover the key card.}
  \Description{Interior view of a medium-sized room with several large crates side-by-side. There is an opening between two of the crates, and an open doorway is at the far end of the room. The player is in front of this open doorway. There are red, yellow, and green boxes drawn on the ground, and a purple line extending from the player.}
  \label{fig:teaser}
\end{teaserfigure}

\maketitle

\section{Introduction}
\input{sec01-intro}

%%%%%%%%%%%%%%%%%%%%%%%%%%%%%%%%%%%%

\section{Related Work}
\label{section:related-work}
\input{sec02-rw}

%%%%%%%%%%%%%%%%%%%%%%%%%%%

\section{Formative Interviews}
\label{section:interview-ov}
\input{sec03-interviews}

%%%%%%%%%%%%%%%%%%%%%%%%%%%

\input{sec04-05-tool-desc}

%%%%%%%%%%%%%%%%%%%%%%%%%%%

\section{User Study}
\input{sec06-user-study}

%%%%%%%%%%%%%%%%%%%%%%%%%%%

\section{User Study Findings}
\input{sec07-findings}

%%%%%%%%%%%%%%%%%%%%%%%%%%%

\section{Discussion}
\input{sec08-discussion}

%%%%%%%%%%%%%%%%%%%%%%%%%%%

\section{Limitations}
\input{sec09-limitations}

%%%%%%%%%%%%%%%%%%%%%%%%%%%

\section{Conclusion}
\input{sec10-conclusion}

%%%%%%%%%%%%%%%%%%%%%%%%%%%


\begin{acks}
We would like to thank Mingyi Li and Michael Malcolm for assisting us with our quality assurance and experimental pilot tests. We would also like to extend our gratitude toward our study participants for their enthusiasm and engagement as well as to the anonymous reviewers for their valuable feedback.
\end{acks}

\balance
\bibliographystyle{ACM-Reference-Format}
\bibliography{main}

% \appendix

\end{document}

%% file: sec01-intro.tex
Mainstream video games have become increasingly accessible to blind and low vision (BLV) audiences in recent years~\cite{Bunting2023}. However, despite these advancements, many accessibility tools introduced within these games diminish the overall fun that BLV players experience and fail to provide them with an experience equivalent to that of sighted players~\cite{Nair2022a, Archambault2008, Porter2013}.

Prior work within game design has emphasized the importance of cultivating specific qualities within games to enhance fun and enjoyment~\cite{Hunicke2004, Garneau2001, LeBlanc2008}. One such quality is the sense of \textit{discovery}, which typically involves players exploring environments for hidden items. Such discoveries are rewarding for players, and these feelings often play a central role in one's experience within a game. Sighted players experience discovery by visually scanning environments and keeping tracking of where they have and have not yet looked --- a process that virtually guarantees that they will discover new items and places.

Assistive tools designed to aid BLV players in navigating game environments, however, often fall short in granting this sense of discovery. In many games, for example, the presentation of worlds is reduced to simplistic lists and grids and players are guided directly from landmark to landmark. This reduces the experience of game completion to a mere point-and-click task, robbing players of the experience of discovering items and places. Indeed, BLV gamers often view such railroading negatively~\cite{Andrade2019, Nair2021}.

A crucial challenge lies in developing in-game navigation assistance tools that \textit{also} foster a sense of exploration and discovery --- providing BLV players an experience on par with that of their sighted counterparts. Thus, in this work, we propose the concept of \textit{exploration assistance} tools within games and explore what such tools for BLV players should offer to facilitate discovery. We use findings from a formative study to create and evaluate an exploration assistance tool, assessing its effectiveness against the navigation tools found within existing accessible games.

We conducted the formative interviews with two experienced BLV gamers, shedding light on their preferences and needs. These interviews revealed that BLV players highly value the experience of traversing game environments through walking, as opposed to relying solely on tools like maps and menus for navigation. They also expressed a desire for certain abilities, such as the capacity to keep track of explored and unexplored areas, discover items while walking through the environment, use audio beacons to quickly navigate to specific locations, and provide multiple approaches for accomplishing navigation tasks within an environment.

In response to the formative interview findings, we created \textit{Surveyor}, an exploration assistance tool for virtual worlds designed to enhance discovery in BLV players. Surveyor includes features such as an exploration tracking system that highlights unexplored areas, the use of the right stick as a radar-like utility so players can scan their surroundings for objects, and the use of menus so players can rapidly query their current exploration status. Figure 1 shows an illustration of Surveyor within a video game environment.

To evaluate Surveyor's effectiveness in promoting the sense of discovery compared to existing accessibility tools, we created a 3D video game and implemented Surveyor alongside two other tools: a simple audio menu (representing tools within audio-based games created for BLV audiences) and a shockwave utility (representing mainstream video game accessibility tools). Our evaluations with nine BLV participants showed that, despite a learning curve, Surveyor granted players with a strong sense of agency in exploring environments and discovering items. The audio menu's simplicity, by contrast, made it preferable for quickly completing levels but diminished the element of surprise in item discovery. The shockwave utility failed to provide players with a meaningful understanding of or immersion within the environment. 

Our findings yield several implications for future exploration assistance tools for BLV players, including that future tools may explore incorporating "hints" into the exploration process and that these tools may also consider granting players the option to "reveal" environments for an easier experience. We also explore the potential of Surveyor-like tools within other contexts.

In summary, we contribute the following within this work:
\begin{enumerate}
    \item a formative study investigating the experiences and abilities important for BLV players to explore game environments and experience a sense of discovery,
    \item Surveyor, an exploration assistance tool for virtual worlds designed to facilitate a greater sense of discovery for BLV players, and 
    \item a user study with nine BLV participants evaluating Surveyor's effectiveness in facilitating discovery against existing in-game accessibility tools.
\end{enumerate}

%% file: sec02-rw.tex
Our work draws inspiration from game design theory, particularly emphasizing the role of discovery in fostering fun within games. We also explore efforts to make mainstream games more accessible to blind and low vision (BLV) audiences and examine work that has looked at exploration within physical environments.

%%%

\subsection{Fun and Discovery in Game Design Theory}
\label{subsec:gdtheory}

Prior work in game design theory has sought to formalize what makes a game "fun." Such a formalization serves a dual purpose --- helping researchers and developers gain a deeper understanding of how existing games facilitate fun experiences while enabling the design of game experiences that more effectively promote a sense of fun in players~\cite{RiotGames2019, Garneau2001}. Perhaps the most well-known framework for formalizing the notion of fun is Hunicke et al.'s MDA ("Mechanics, Dynamics, and Aesthetics") Framework, which introduces the "8 Kinds of Fun"~\cite{Hunicke2004}. These eight categories --- which encompass sensation, fantasy, narrative, challenge, fellowship, discovery, expression, and submission --- manifest at varying levels across different games and in-game experiences. In this work, we look specifically at the element of \textit{discovery}, a type of fun that sees games as "uncharted territory" to be explored and uncovered by players, and prior work has found this exploration to be a major part of one's gaming experience~\cite{Cole2021}. In this work, we explore how to build gaming experiences tailored toward BLV players that better facilitate a sense of discovery and, thus, fun.

%%%

\subsection{Fun and Autonomy in Games for BLV Audiences}

Prior work has found that games designed specifically for BLV players often lack the same level of enjoyment and complexity as mainstream games~\cite{Smith2018, Andrade2019, Goncalves2020a}. One of the primary reasons for this is the substantial simplification employed by these games. For example, game worlds are typically presented to players as basic lists and grids~\cite{Trewin2008, Westin2004, Matsuo2016} and their movement through the environment is heavily guided~\cite{NaughtyDog2020, NaughtyDog2022}, leaving little room for independent exploration and the discovery of hidden objects. This deprives BLV players of a sense of autonomy in exploring game environments and uncovering hidden items and places on their own terms.

Many mainstream games are inaccessible to BLV players~\cite{Porter2013}, yet a significant number of BLV players express a strong desire to engage with these games. Work by Gon\c{c}alves et al., for example, revealed how BLV gamers adapt by using existing features within these games, such as sound design, to overcome their general inaccessibility~\cite{Goncalves2023b}. Despite their efforts, BLV players still face challenges due to missing information and the increased cognitive effort needed to play these games without the necessary assistive tools, resulting in a less enjoyable experience.

Issues with current games point to a need for in-game accessibility tools that better promote a sense of \textit{autonomy} in BLV players. Prior work has found that, within games, a sense of autonomy is linked to motivation and well-being in players~\cite{Tyack2021}, leading to increased enjoyment and the desire to further play through the game~\cite{Ryan2006}. Although these particular studies were performed with sighted players, BLV players, too, have negatively viewed the railroading present in many games targeted toward them~\cite{Andrade2019, Nair2021}. In this work, we build tools to further promote a sense of autonomy in BLV players, and we do this by better facilitating a sense of discovery --- allowing for more enjoyable gaming experiences.

\subsection{Game Accessibility Tools for BLV Players}

Gaming for BLV audiences has largely centered around "audio games" --- games that present audio-based experiences and which often feature little to no visuals~\cite{Friberg2004, Nair2021}. Many well-known audio game titles --- such as \textit{Terraformers}~\cite{Westin2004, PinInteractive2003}, \textit{A Hero's Call}~\cite{OutOfSightGames2017}, and \textit{ShadowRine}~\cite{Matsuo2016} --- represent environments as lists and two-dimensional grids~\cite{Goncalves2023b}. However, this approach simplifies worlds that may have otherwise been rich and highly detailed were they designed for sighted players --- thus resulting in inequivalent gaming experiences~\cite{Andrade2019}. Some audio games attempt to work around this by integrating audio cues into the environment to enhance players' environmental awareness~\cite{Nair2021}. Despite this, games created for BLV audiences are still not comparable to mainstream games targeted toward sighted players in terms of complexity and fun~\cite{Andrade2019, Nair2021, Goncalves2023b}.

Although mainstream games have been largely inaccessible to BLV players, game developers have begun introducing tools tailored to their needs. Notably, \textit{The Last of Us Part 2} (2020)~\cite{NaughtyDog2020} and \textit{The Last of Us Part 1} (2022)~\cite{NaughtyDog2022}, created by Naughty Dog Entertainment, have incorporated features like an echolocation-inspired "enhanced listen mode" --- which allows BLV players to perceive objects in the game environment --- alongside assisted traversal and combat mechanics~\cite{Gallant2020, McAllister2022}. The enhanced listen mode often spoils the locations of hidden objects and enemies, however, and traversing the environment is a highly automated process.

The academic community has also recently developed tools for making 3D video games more accessible to BLV players. Nair et al., for example, introduced the NavStick system to provide BLV players with the means to "look around" themselves in an accessible manner~\cite{Nair2020a, Nair2021}. Their evaluations of NavStick showed that it offered BLV players an enhanced sense of agency when compared against tools in existing BLV-accessible games. However, their work pointed out the need for tools that provide broader overviews of game worlds beyond just the player's immediate area --- overviews that can facilitate more thorough explorations of a game world. 

Our present work on fostering discovery for BLV players builds upon these previous works by investigating how to give BLV players the ability to obtain a richer overview of their surroundings while not spoiling the locations of items around them. 

%%%

\subsection{Facilitating Exploration for BLV Users in Physical Environments}

Work looking at how BLV users can better explore environments has also extended to the physical realm. Prior work within physical~\cite{Banovic2013, Giudice2018, Hill1993} and virtual~\cite{Connors2014, Lahav2008a} contexts has emphasized the importance of exploring environments so that BLV people can form cognitive maps and better engage with their environments. This has inspired further work looking at how assistive tools can help users better explore spaces.

Clemenson et al., for example, argued that navigation systems within the physical world should be rethought so that users can better engage with the world~\cite{Clemenson2021}. Jain et al. looked specifically at how navigation assistance tools for BLV users can better assist them in exploring the environments they are traveling through, and they present design recommendations for creating exploration-oriented navigation systems~\cite{Jain2023}. Some existing navigation tools also offer basic features to allow BLV users to explore their environments. NavCog3, for instance, announces nearby points of interest as a user is traveling a pre-defined route~\cite{Sato2017}, and the now-defunct Microsoft Soundscape used spatialized sound to notify users of nearby landmarks~\cite{MicrosoftResearch2018}. Little work, however, has been done to investigate how assistive navigation tools can be rethought to help BLV users better explore video game environments and evaluate the role that fun plays in the exploration process.

%% file: sec03-interviews.tex
We conducted fully remote formative interviews with two BLV gamers to gain insights into the experiences and abilities that are important for them for exploring game environments and experiencing a sense of discovery. We recruited both participants (referred to as P1 and P2) from a mailing list of BLV individuals who participated in our laboratory’s previous studies. P1 was aged 26-35, and P2 was aged 36-45. Both participants were male, described themselves to have total blindness, and stated that their vision impairments developed when they were very young (P1 became totally blind at around four years old, and P2 at birth). Both participants reported themselves as being very experienced with video and audio games (4+ on a 5-point Likert scale).

We conducted two-part, semi-structured interviews with both participants. In the first part, we asked participants to discuss video and audio games they were familiar with that featured environments requiring navigation and exploration. For each game they mentioned, we asked about the specific features within those games that facilitated their navigation through the environments. We also gathered their opinions on what aspects they liked and disliked about these features. In the second part, we presented hypothetical game scenarios that involved exploring various game worlds, and we encouraged participants to share their thoughts on the features and tools they need in order to effectively navigate and explore these environments. Each interview lasted no more than one hour.

We followed an inductive coding process that involved two members of the research team to analyze these interviews. Each coder went through study session transcripts and coded quotes. Then, both coders iterated on the codes together until they could not iterate further. The study was approved by our Institutional Review Board, and sessions were recorded with participants' consent.

\subsection{Abilities Desired by BLV Users to Effectively Explore Game Environments}

P1 and P2’s sentiments revealed four major abilities that BLV users desire in order to effectively explore game environments. These four abilities will serve as design goals that exploration assistance systems for BLV players --- such as Surveyor --- must address in order to facilitate exploration and, thus, a sense of discovery for BLV players within games.

\subsubsection{\textbf{Design Goal 1}: Discovering items while physically walking through the world.}

Both participants expressed a strong preference for physically walking through game environments rather than relying solely on menus or maps to discover items and landmarks. To illustrate this point, both P1 and P2 independently mentioned a feature in the game \textit{A Hero's Call}~\cite{OutOfSightGames2017}, a well-known title in the audio game community, which provides a grid-based top-down map to players. This map allows players to navigate and locate landmarks without encountering obstacles in the game world. However, both participants reported that they rarely used the map feature, as they found the act of physically exploring the game's environment to be more enjoyable. P1 specifically emphasized that walking around the environment provided a heightened sense of immersion:

\begin{quote}
    \textit{"Honestly, I prefer being on the ground moving with all the sounds around me and everything. There's just more cues to experience, and it's more fun that way."} --- \textbf{P1}
\end{quote}

In a similar vein, P2 mentioned how "doing the work" to look for objects provided them with a greater sense of accomplishment when they find objects that they did not expect to be there --- indicating that feelings of discovery are a crucial part of facilitating fun within a game. As per this sentiment, exploration assistance systems within games should facilitate players' immersion within the environment --- allowing them to move through it and discover items on their own accord.

\subsubsection{\textbf{Design Goal 2}: Keeping tabs on explored and unexplored regions within the world.}
\label{subsec:formative-dg2}

Both P1 and P2 desired to track their exploration progress by keeping tabs on visited and unvisited areas within the game environment. When discussing a hypothetical game scenario featuring a dungeon, P2 drew a parallel between this desired ability and the natural abilities of people with sight:

\begin{quote}
    \textit{"Letting someone know if they have or haven't been in a space is great, because again, you [as a sighted person] can go into a room and be like, 'Oh yeah, this is the room with the skeleton' or 'This is the room with the really cool torches.' [...] But as a blind user, we don't really have the bandwidth or info to know that."} --- \textbf{P2}
\end{quote}

P2's thoughts highlight the significance of being able to "recognize" areas and objects as an important ability for enhancing their sense of fun and discovery within a game world. Exploration assistance systems should facilitate this recognition by giving players the ability to identify what they have seen so far.

% highlighting what players have and have not yet seen.

\subsubsection{\textbf{Design Goal 3}: Using audio beacons to quickly navigate to specific locations.}

Both P1 and P2 mentioned that audio beacon functionality was crucial for quickly locating and navigating to points and objects without having to go through the process of actively re-exploring the environment in an attempt to find it again. P2, in particular, made an analogy with the physical world by bringing up audio beacon features within Microsoft's Seeing AI~\cite{Microsoft2017}:

\begin{quote}
    \textit{"Being able to put beacons on things --- tag them --- is always helpful. Speaking of which, Seeing AI just dropped an updated 'world channel,' [...] and one of the things that you can do with the LIDAR is you can scan around a space and then place beacons on items to get an idea of where they're at. So. I'm a big fan of 'beacon-ing' things, because it allows us to have real points we can track in space."} --- \textbf{P2}
\end{quote}

Beacons can save time for BLV users looking for a specific target, both within games and the physical world. However, within games, P1 warned that using beacons too much can make games too easy:

\begin{quote}
    \textit{"I feel like beacons are more like a thing that you want to use in a game where the goal is to stop the player from getting lost or confused when trying to find something. But you can't make it too easy on them, right? Otherwise the challenge is gone and they don't want to play it anymore."} --- \textbf{P1}
\end{quote}

The importance of beacons to both P1 and P2 indicates that an exploration assistance tool for games should also provide audio beacon functionality. However, it must still preserve an element of challenge by not make finding points of interest too easy.

\subsubsection{\textbf{Design Goal 4}: Providing multiple approaches for understanding and navigating through the environment.}

Both participants emphasized the importance of having multiple ways to understand the contents and general characteristics of the environment and navigating through the environment (e.g., finding an object or making their way to a specified destination). P2, in particular, praised \textit{A Hero's Call}~\cite{OutOfSightGames2017} for providing multiple tools to players to understand the environment:

\begin{quote}
    \textit{There's no one-size-fits-all approach. [...] I'm a big believer in the more tools that you can give somebody to understand the world, the better it is for the player. That's why I like \textit{A Hero's Call} as much as I do, because they went: 'Let's do a map,' 'let's do beacons,' 'let's do trackable things,' 'let's do coordinates.' Like, if you want it, it's there for you.} --- \textbf{P2}
\end{quote}

P2's sentiments emphasize the importance of choice while exploring and making sense of a game environment, especially in terms of the \textit{actions} players can take in order to achieve their goals. P1, however, also talked about facilitating multiple approaches to moving through the environment as well:

\begin{quote}
    \textit{"I guess you could just use beacons to move through the level, but I would prefer to do it manually because I still want some challenge. [...] If I get lost or something, then I'll use beacons.} --- \textbf{P1}
\end{quote}

Both participants' sentiments indicate that exploration assistance tools for games should also facilitate a high level of player choice and autonomy as they understand and move through a game level.

%% file: sec04-05-tool-desc.tex
\section{Surveyor}
\label{sec:surveyor-desc}

Using the insights from our formative study, we created Surveyor, an exploration assistance tool for virtual worlds designed to facilitate a greater sense of discovery for BLV players. Figure~\ref{fig:teaser} shows an illustration of Surveyor within our game testbed. The accompanying demo video also demonstrates Surveyor's functionality.

Surveyor comprises two core interactive components: a right thumbstick-based scanning tool and an audio-based menu system. These components work on top of a matrix-based data structure that tracks regions within the current area that the player has and has not yet explored. In the following subsections, we describe the design and implementation of Surveyor in detail and explain how Surveyor satisfies the design goals from our formative interviews. 

We implemented Surveyor using the Unity game engine~\cite{UnityTechnologies2021}. All interactions with Surveyor are performed using a standard game controller (e.g., an Xbox or PlayStation controller).

\subsection{Right Stick-Based Directional Scanning}

Surveyor's first core component is a scanning utility that leverages the right stick on a game controller so that BLV players can ``look around'' themselves. When the player tilts the right stick in any direction, the game will announce the name of the first item that lies in that direction within their current line-of-sight. Players can use the right stick to survey in any direction. The information that the game returns updates in real-time as the player moves through the environment, meaning that players are free to move through the world as they wish. This utility satisfies \textbf{Design Goal 1} (discovering items while physically walking through the world).

We adopted this utility from prior work that has built and investigated direction-based scanning in video game environments (e.g., NavStick ~\cite{Nair2020a, Nair2021}) as well as in the physical world (e.g., Talking Points 3~\cite{Yang2011}). Within games specifically, research has shown that directional scanning allows BLV players to obtain a quick overview of their surroundings and better understand their environment~\cite{Nair2021, Nair2022a}, and so we integrated this utility to allow players to quickly get a sense of their immediate surroundings.

\subsection{Matrix-Based Exploration Tracking}

\begin{figure*}
    \centering
    \includegraphics[width=0.99\textwidth]{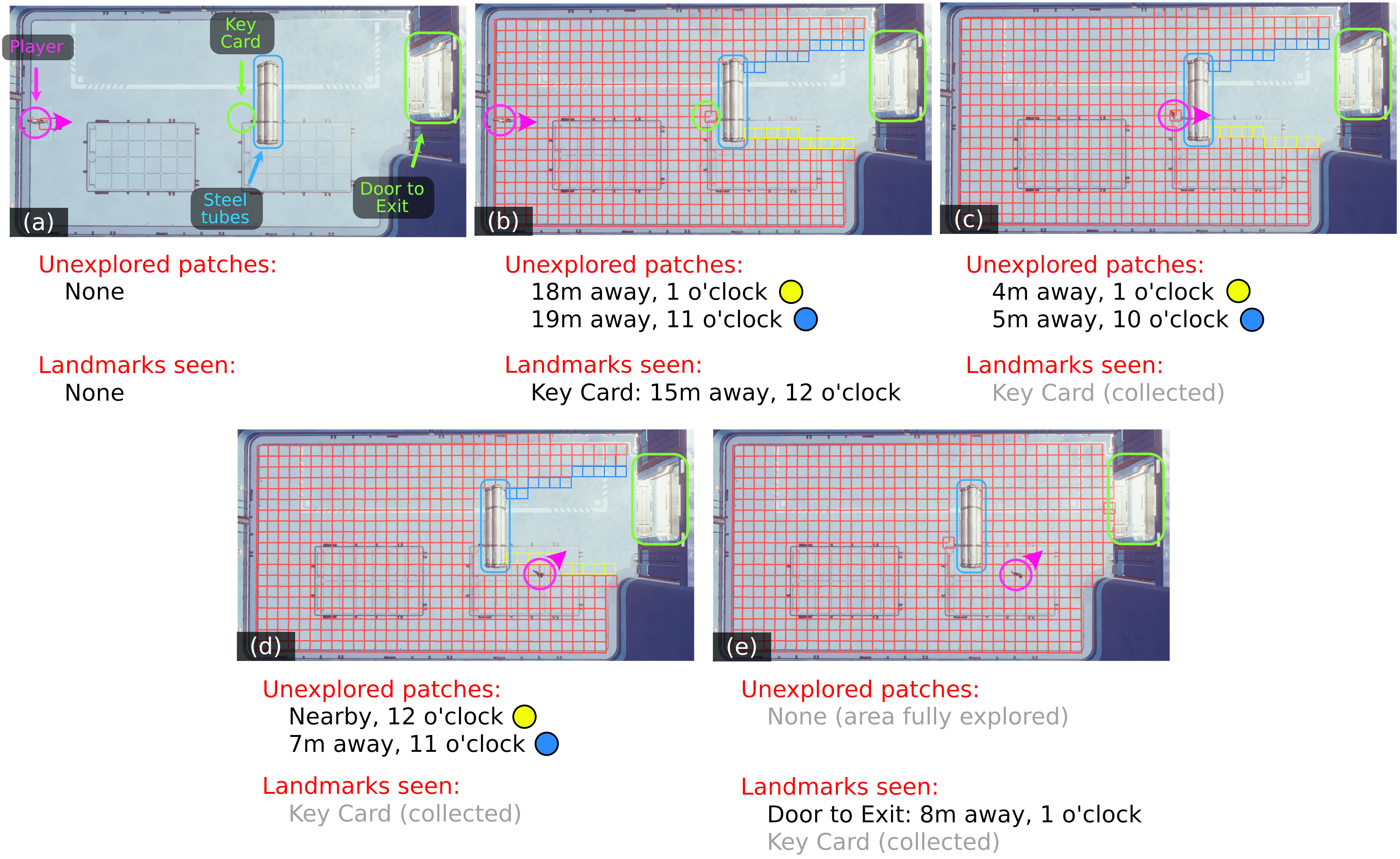}
    \caption{Surveyor's matrix-based exploration system in a room within our game. Text below each map shows unexplored patches and landmarks at that time point. Starting from the left side, the player aims to reach the unseen door on the right, initially obstructed by the large steel tubes in the center. A key card lies in front of the tubes. The player begins the level (A), surveys their line-of-sight with the right stick (B), and discovers the key card and two unexplored "patches" of cells \textit{behind} the tubes. After collecting the key card (C), the player moves to the nearest patch (D), surveys again (E), and discovers the exit door.}
    \centering
    \Description{Five copies of a top-down view of a somewhat empty room. There are boxes drawn on the floor in some of these view representing the exploration status of the room in a given panel. There are steel tubes situated horizontally on the ground in the middle of the room obstructing the player's initial view of the door which lies several meters behind these tubes.}
    \label{fig:gridprog}
\end{figure*}

Directional scanning, while valuable, is limited in that it does not provide any information about the greater world beyond one's line-of-sight, which is crucial to be able to explore it. To overcome this limitation, we developed a systematic approach to track explored and unexplored regions within the current area. Figure~\ref{fig:gridprog} illustrates this approach.

Each room within the world is represented internally using a matrix, with each \textit{matrix cell} assigned a specific state. Cells traversed by the player or covered during a sweep of the right stick are labeled as \textit{"explored."} If the scanning utility encounters an obstruction, such as a wall or large object, cells corresponding to the obstruction will be marked as \textit{"obstructions."} Areas outside of these cells remain unknown to Surveyor and, consequently, the player.

As the player explores the area --- either via right-stick scanning or physical movement --- some parts of the area will consist of "explored" matrix cells while other parts remain unknown. These unknown regions (also known as "unexplored patches") typically lie behind large obstructions or around corners --- areas beyond the reach of the right-stick scanning utility, which operates exclusively via line-of-sight. (Note that this matrix is solely used as Surveyor's underlying data structure and is not exposed to the user.)

Surveyor's primary objective is to highlight these unexplored patches, enabling players to survey them using the right stick should they choose to do so. In practice, Surveyor conveys information about the boundaries separating "explored" cells and unexplored patches via a menu system we describe in the next subsection. These boundaries represent the "cusp of the unknown" for both the player and the system itself. An area is considered fully explored when no unexplored patches remain --- at this point, all of the area's walls have been seen, and all cells within those wall bounds are marked as "explored." This functionality satisfies \textbf{Design Goal 2} (keeping tabs on explored and unexplored regions within the world).

%%%%%%

\subsection{Exploration Status Menu System}

Players can access their current exploration status via Surveyor's audio-based menu system. We integrated a menu system to facilitate easy access to exploration information and since prior work has recommended integrating menus alongside other tools --- such as directional scanning utilities --- to facilitate a more complete sense of spatial awareness than that facilitated by menus alone~\cite{Nair2022a}. The upper face button on the player's controller activates the menu, which consists of four sub-menus:

\begin{enumerate}[leftmargin=14px]
    \item \textbf{Unexplored patches in the player's current area.} This sub-menu lists unexplored patches within the the player's current area (room, corridor, etc.). For each patch, the game announces the distance to the patch as well as its direction relative to the player. Each patch is announced via spatialized sound emanating from the direction of the patch. \textit{(Example announcement: "Unexplored patch, 5 meters to your 3 o'clock.")}
    \item \textbf{Landmarks seen in the player's current area.} This sub-menu catalogs points-of-interest seen by the player within the current area. Landmarks are considered seen when the player tilts the right stick toward the object or moves close enough to the object to hear any associated sound effects. For each landmark, the game announces the player's distance to the object as well as its direction with respect to the player. Each landmark is announced via spatialized sound emanating from the direction of the landmark. \textit{(Example announcement: "Key card, 15 meters to your 7 o'clock.")}
    \item \textbf{Other known unexplored areas within the greater world.} This sub-menu lists known areas within the broader game world that have not yet been fully explored. An area becomes known when the player points the right stick at a door leading to that area. If the player does not enter the area, the sub-menu will announce that the area "is unexplored." If the player has been inside the area but has not fully explored it, the sub-menu will announce the number of unexplored patches within that area. \textit{(Example announcements: "Storage room is unexplored" and "Laboratory, 7 unexplored patches.")}
    \item \textbf{Fully explored areas within the greater world.} This sub-menu lists areas within the world that have been fully explored by the player. As mentioned previously, an area is deemed fully explored when it has no more unexplored patches. Each area is announced simply using its name.
\end{enumerate}

In all sub-menus, players can press the right bumper button to activate a looping audio beacon that will guide them toward the specified element. For the "unexplored patch" and "seen landmarks" sub-menus, the beacon leads the player to the edge of the unexplored patch or to the landmark, respectively. In the "unexplored areas" and "fully explored areas" sub-menus, the beacon directs the player to the entrance of the designated area. The availability of the audio beacon in all sub-menus satisfies \textbf{Design Goal 3} (using audio beacons to quickly navigate to specific locations). 

Players, however, are not required to use beacons for navigation. They can track objects as they move through a room by monitoring their positions using the right stick or by relying on spatialized sound effects that objects in the world may emit. Additionally, players can seamlessly transition between using the right stick to make sense of their immediate environment and using the menus to search for specific elements. This freedom to use whatever tool they want satisfies \textbf{Design Goal 4} (providing multiple approaches for understanding and navigating through the environment).

%%%%%%%%%%%%%%%%%%%%%%%
%%%%%%%%%%%%%%%%%%%%%%%
%%%%%%%%%%%%%%%%%%%%%%%

\section{Status Quo Tools}
\label{sec:status-quo-desc}

In preparation for our user study, we implemented two additional tools alongside Surveyor within our game. These tools --- a simple audio-based menu and a shockwave utility --- represent \textit{existing} approaches for making games accessible to BLV players. In this section, we describe both tools' implementation and operation. The accompanying demo video also shows these tools.

\subsubsection{Simple Audio-Based Menu}

The simple menu represents the use of straightforward audio-based menus for communicating what is in the game world. Many audio-based games targeted toward BLV audiences~\cite{OutOfSightGames2017, PinInteractive2003, Westin2004, Kaldobsky2011} use menu-based representations to present the world to players. Thus, given that menu-based approaches have been widely used in audio games~\cite{Goncalves2023b}, we wanted to directly compare Surveyor with these approaches and capture BLV players' sentiments when they use both tools within the same study session.

Players activate the tool by pressing the left face button on their controller to open a list of points-of-interest within the room they are currently in. As players scroll through the list using the D-pad, they hear each item’s name announced via spatialized sound. Players can select any item within the menu using the right bumper button on their controller to activate an audio beacon that will guide them toward the item --- this is the same audio beacon system utilized by Surveyor. The simple audio menu is modeled after existing games and research in that it employs an alphabetical ordering of items~\cite{Westin2004, PinInteractive2003, Nair2021, Nair2022a}.

\subsubsection{Shockwave Utility}

The shockwave utility emulates the "enhanced listen mode," an accessibility feature within \textit{The Last of Us} games~\cite{NaughtyDog2020, NaughtyDog2022} that received wide recognition and acclaim for making the games accessible to BLV audiences~\cite{Gallant2020, McAllister2022, Molloy2020, Watts2021, Bunting2022, Croft2022} and is inspired by echolocation techniques within the real world~\cite{Nair2022a}. We chose to include the shockwave in our evaluation because it represents mainstream games and tools recently implemented by game developers to open these games up to BLV audiences.

Upon activation, the utility emits an acoustic shockwave starting from the player's position and expanding outward. When the shockwave hits any object within the room, a ping emanates from that object via spatialized sound. If the shockwave strikes a wall, a burst of white noise emanates from that position. The shockwave corresponds to real-world physics in that closer objects will emanate their sounds back to the player before more distant objects.

Similar to \textit{The Last of Us}'s implementation, players can press a button on their controller to orient their character in the direction of level progression after emitting a shockwave. In our own "golden path" implementation, players may be oriented toward key cards, the checkpoint, or doors that will eventually lead players to these landmarks. Players can then activate a looping audio beacon that leads them to these landmarks --- this is the same beacon system that both Surveyor and the simple audio menu use.

Finally, similar to how the "enhanced listen mode" allows players to selectively scan for items or enemies within their immediate area --- thus allowing players to filter what the tool plays back --- the shockwave allows players to filter which types of objects emanate sounds in response to a shockwave. Our implementation offers the choice to filter between "all" items, "objective" items (including key cards, doors, and checkpoints), "decorative" items (comprising all items not classified as "objective" items or walls), and "walls."

%% file: sec06-user-study.tex
We performed a user study to address two research questions:

\begin{enumerate}%[label=(RQ\arabic*), leftmargin=2.5\parindent]
    \item[(RQ1)] How well does Surveyor facilitate a sense of exploration and discovery for BLV players within 3D game environments?
    \item[(RQ2)] In what ways does a player's experience with Surveyor differ from that with navigation assistance tools within existing accessible games?
\end{enumerate}

We integrated Surveyor, along with the two ``status quo tools'' --- the simple audio-based menu and the shockwave utility --- into a 3D adventure video game we created called \textit{Cosmic}. Our user study was fully remote and involved nine BLV participants. 

Before executing our studies, we performed preliminary tests to ensure that we presented the best possible version of our three tools to our study participants. We conducted these with a representative participant (a BLV person who was a member of our laboratory during the project) and a non-representative participant (a sighted but blindfolded person who was not part of the project). We first held a \textit{quality assurance (QA) session} with our non-representative participant to identify critical bugs within the study game, all three tools, and study script that could derail a study session or otherwise compromise the quality of our BLV pilot tester's feedback. We note that sighted but blindfolded testers are typically not suitable proxies for BLV testers~\cite{Sears2012, Silverman2015}, and as such, our actions following this session were limited to bug fixing.

After we fixed issues identified during our QA session, we performed a full \textit{experimental pilot session} with our representative (BLV) participant. This BLV user provided valuable feedback oriented toward the design of the game and the tools. Fixes following this pilot included truncating some of Surveyor's announcements, tweaking Surveyor's menu presentation, and modifying other audio-based aspects of the game world. (Note that our BLV pilot tester was selected only for these pilot tests and did not participate in our formative interviews or main study.)

The following subsections detail our study game testbed, study conditions, participants, and study procedure.

\subsection{Game Testbed}

In \textit{Cosmic}, players assume the role of a spacefaring explorer who has just landed in an abandoned space station. To escape, players must traverse the station's various levels, which we deliberately designed to feature complex environments. These environments include obstructions necessitating players to move around corners to explore new areas, as well as numerous decorative objects, allowing us to observe how players explore the environment using the provided tools.

We created \textit{Cosmic} using the Unity game engine~\cite{UnityTechnologies2021}. We designed the level layouts using the Dungeon Architect Unity asset~\cite{DungeonArchitect}, which ensured level layouts that were equivalent in difficulty by defining a "grid flow" that defined the basic high-level structure of the level. Figures ~\ref{fig:teaser} and ~\ref{fig:gridprog} show in-game views from \textit{Cosmic}.

\begin{figure*}
    \centering
    \includegraphics[width=0.9\textwidth]{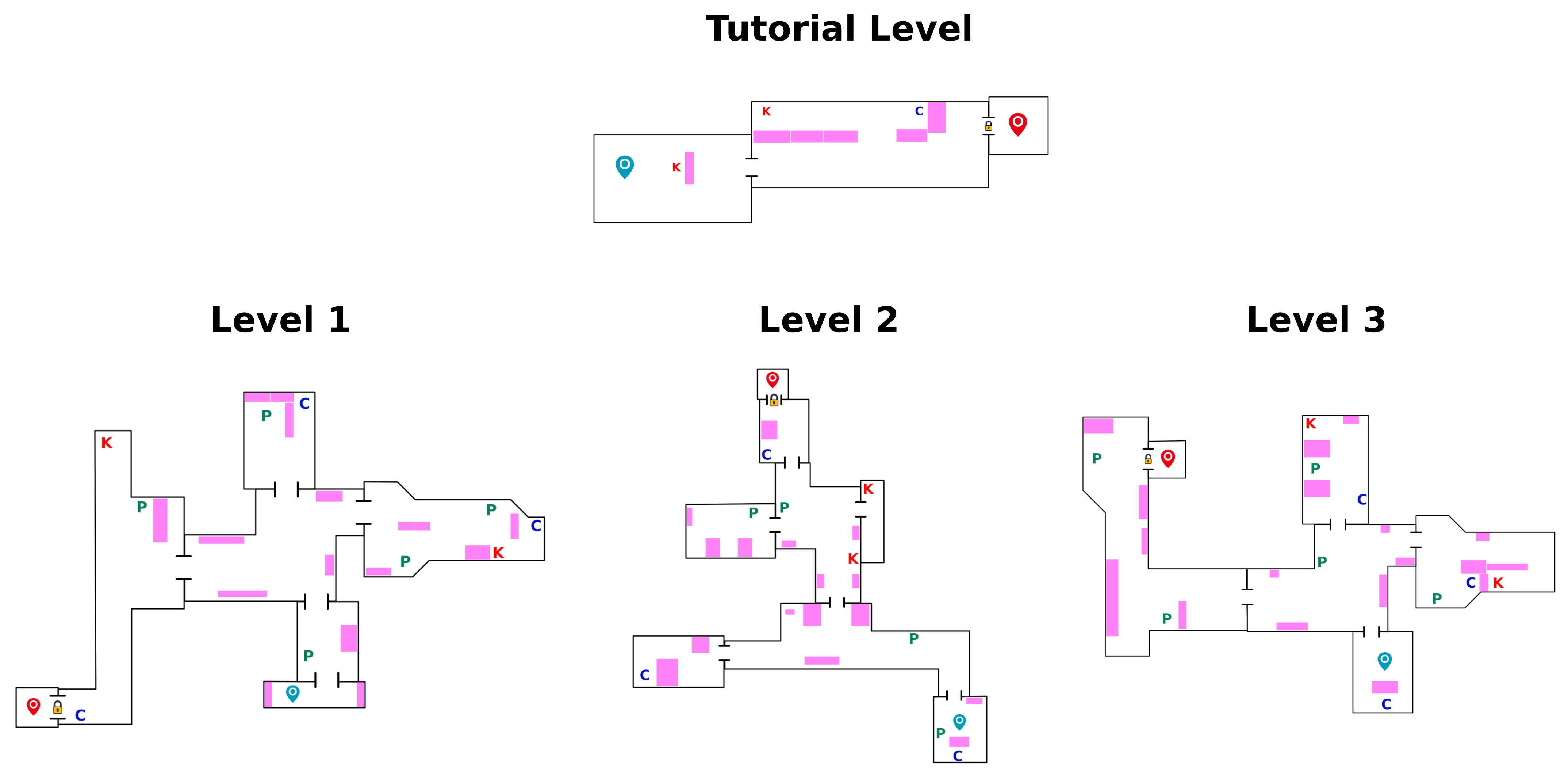}
    \caption{Maps of tutorial and main levels. Players navigate from start (\textcolor{MapKey_PinBlue}{\textbf{blue}} map pin) to checkpoint (\textcolor{red}{\textbf{red}} map pin). The checkpoint room is initially locked (lock symbol). Players must collect two key cards (\textcolor{red}{\textbf{K}}) to unlock the door. All levels contain multiple decorative objects which are announced by the three study tools. \textit{Decorative objects}: \textcolor{MapKey_Pink}{\textbf{Pink}} boxes = physically obstructive objects; \textcolor{MapKey_Green}{\textbf{P}} = non-obstructive (point) objects; \textcolor{MapKey_Blue}{\textbf{C}} = log entry consoles.}
    \Description{Floor plan views of the game's levels. Each view contains multiple boxes, markers, and symbols indicating the positions of points-of-interest within the level.}
    \label{fig:leveloverhead}
\end{figure*}

The game comprises three main levels with distinct layouts and a shorter tutorial level intended to acquaint players with each tool. Figure \ref{fig:leveloverhead} shows maps of the tutorial level and the main levels. In each level, players must find and collect two key cards, which allow access to a final checkpoint past a locked door. Areas throughout the levels contain decorative objects, such as crates, barrels, and "log entry consoles" that play voiced clips revealing the abandoned space station's backstory. Players start each level with no knowledge of level layouts or object positions --- they only know that they have to collect two key cards to reach a checkpoint. 

In our game, players control the character using the left thumbstick. Pushing it forward or backward moves the character in those directions, while tilting it left or right rotates the character. This control scheme mirrors controls commonly found in mainstream 3D games that employ a fixed over-the-shoulder camera perspective~\cite{TombRaider, MetroidPrime1, MetroidPrime2, MetroidPrime3, SilentHill}, utilizing left/right inputs on the left thumbstick for character rotation.

To enhance player awareness, we introduced various audio cues. When a player physically collides with an obstruction, they will hear a scraping sound emanating from the point of impact. Objects (such as keys, obstacles, and checkpoints) emit relevant sounds when the player approaches within four meters. Additionally, entering a new area triggers an announcement of the area's name (e.g., "Laboratory" or "Office"). Players can manually trigger this announcement whenever they wish by pressing the right face button. These cues help BLV players stay informed about crucial in-game events, such as wall collisions and room transitions --- events that sighted players can perceive visually.

%%%

% insert table here
\input{sec06_TABLE-ptcpts-demographics}

%%%

\subsection{Participants}
\label{subsec:us-ptcpts}

Our nine BLV participants included the two participants from the formative interviews (referred to, again, as P1 \& P2 in this study) whom we invited back to test our new tools. We recruited the other seven from our laboratory's mailing list and through posts on the AudioGames.net Forum\footnote{\url{https://forum.audiogames.net/}}. Table \ref{tab:demographics} shows demographic information for all nine participants. Two participants (P1 \& P2) reported mild-to-moderate hearing loss in one of their ears. 

Eight of nine participants reported themselves as being very or extremely experienced with video and audio games (4+ on a 5-point Likert scale); one participant (P3) described themselves as being moderately experienced (3 on a 5-point scale). All nine participants reported experience with audio games — games mentioned by multiple participants include \textit{A Hero's Call}~\cite{OutOfSightGames2017}, \textit{Terraformers}~\cite{PinInteractive2003}, \textit{ShadowRine}~\cite{Matsuo2016}, \textit{The Vale}~\cite{FallingSquirrel2021}, and \textit{A Blind Legend}~\cite{Dowino2016}. Six participants reported playing mainstream fighting games, such as \textit{Mortal Kombat}~\cite{Midway1992}, which are largely playable by BLV players~\cite{Andrade2019}. Five participants also reported experience with one or both of \textit{The Last of Us} games~\cite{NaughtyDog2020, NaughtyDog2022}.

\subsection{Procedure}

We began the session by administering a pre-study questionnaire requesting demographic information. After the pre-study questionnaire, we provided participants with a general overview of the game, which included a description of movement controls.

We then moved onto the first tool. For each of the three tools, we followed an identical procedure:

\begin{enumerate}
    \item We read out a description of the tool's features and controls.
    \item Participants practiced using the tool in the tutorial level, which was the same for all three tools.
    \item After addressing any final questions, participants used the tool to complete one of the three main levels, each associated with a different tool.
    \item When participants reached the level's checkpoint, we administered a post-level questionnaire gauging participants' opinions on the tool itself especially with respect to how well it helped them explore the level.
\end{enumerate}

Participants used each tool within one of the three main levels. Although all participants played the three levels (which were similar in length and difficulty) in the same order, we counterbalanced the order of the tools via a Latin square design to minimize order-related effects. At the end of the study, we administered a post-study questionnaire, which included a forced ranking of the tools.

The technical setup for this study involved distributing a pre-built Windows executable of our game. We signed each executable with a unique identifier for each participant. Using this identifier, the game communicated with a cloud-based backend, allowing us to control the game's state (current level, active tool, etc.) remotely and in real time.

\subsection{Data Collection and Analysis}

We administered all questionnaires by having the facilitator read out each question and input the participant’s response into an internal Google Form. We recorded all sessions with participants’ permission for transcription purposes. We also obtained raw data of participants’ actions within the game by capturing logs via our cloud backend. We followed an inductive coding process identical to that of the formative interviews to identify themes in participants' open-ended sentiments. This involved two members of the research team who individually went through study session transcripts and coded quotes. Afterwards, both researchers iterated on the codes together until they could not iterate further. This study was approved by our Institutional Review Board, and sessions were recorded with participants' consent. Sessions lasted no more than 90 minutes.

%% file: sec06_TABLE-ptcpts-demographics.tex
\begin{table*}[t]
\begin{tabular}{c|c|c|l|l|l}
\textbf{ID} & \textbf{Gender} & \textbf{Age}   & \textbf{Vision Impairment (VI)}                                                                           & \textbf{VI Onset}                                                                                                   & \textbf{Hearing Impairment}         \\ \hline \hline
P1 & M      & 26-35 & Total blindness                                                                                  & \begin{tabular}[c]{@{}l@{}}Low vision at birth;\\ total blindness at $\sim$4yrs\end{tabular} & Mild loss in left ear          \\
P2 & M      & 36-45 & Total blindness                                                                                  & At birth                                                                                                   & Moderate loss in right ear \\
P3 & M      & 18-25 & Total blindness                                                                                  & At birth                                                                                                   & None                       \\
P4 & M      & 18-25 & Total blindness                                                                                  & At birth                                                                                                   & None                       \\
P5 & M      & 36-45 & \begin{tabular}[c]{@{}l@{}}Total blindness in one eye;\\ 20/800 acuity in other eye\end{tabular} & \begin{tabular}[c]{@{}l@{}}One eye non-functional at birth;\\ other eye gradually lost vision\end{tabular} & None                \\
P6 & M      & 18-25 & Total blindness                                                                                  & At birth                                                                                                   & None                       \\
P7 & M      & 18-25 & Light perception only                                                                            & At birth                                                                                                   & None                       \\
P8 & M      & 26-35 & Light perception only                                                                            & At birth                                                                                                   & None                       \\
P9 & M      & 18-25 & Total blindness                                                                                  & At birth                                                                                                   & None                      
\end{tabular}
\vspace{2mm}
\caption{Demographic information of all nine main study participants.}
\label{tab:demographics}
\end{table*}

%% file: sec07-findings.tex
In this section, we report the main takeaways from our user study, where we look at how well Surveyor facilitates a sense of exploration and discovery for BLV players and how players' experiences with the tool compare to that with existing techniques --- in this case, a menu-based system and a shockwave utility inspired by \textit{The Last of Us} games.

\begin{figure}
    \centering
    \includegraphics[width=0.45\textwidth]{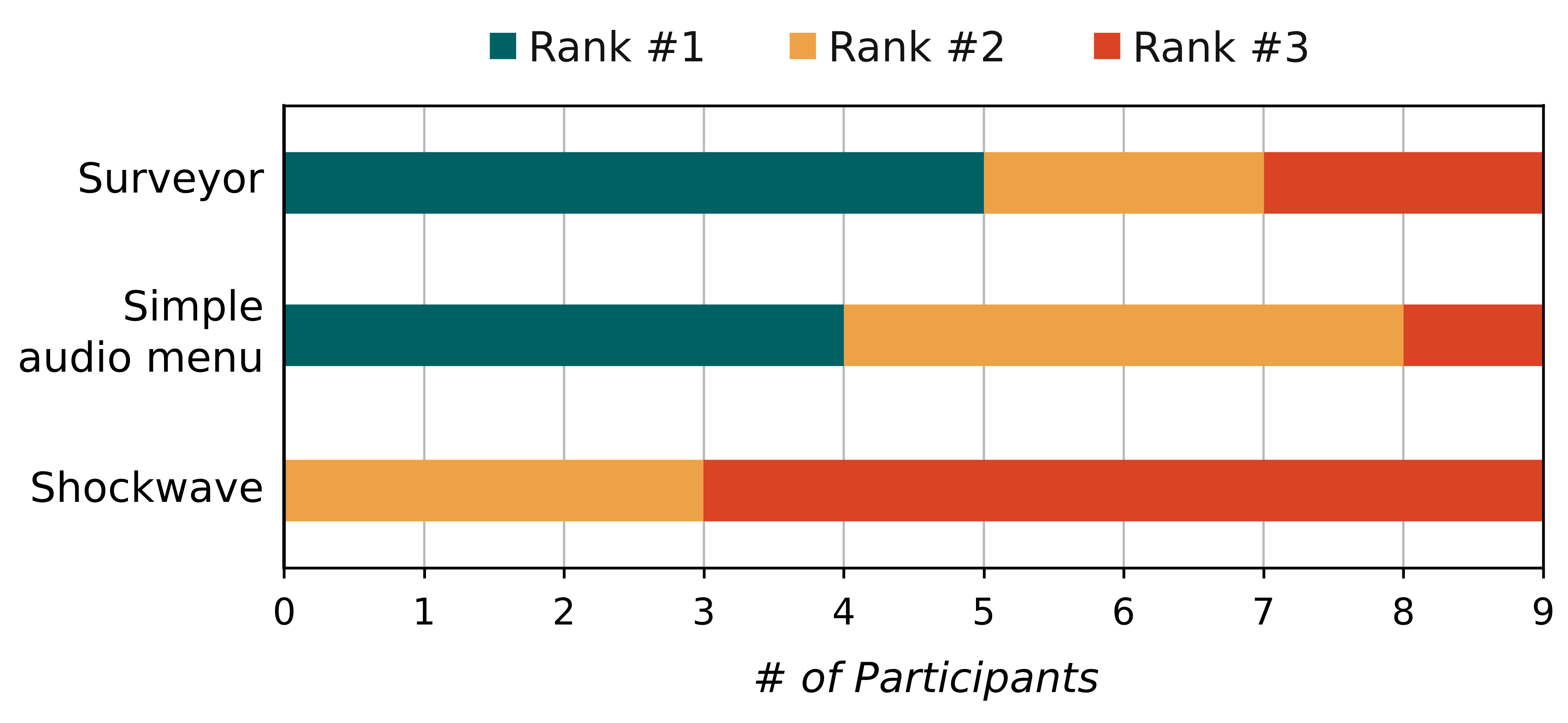}
    \caption{Forced ranking results from post-study. Five participants ranked Surveyor as their favorite tool while four participants ranked the simple audio menu as their favorite. No participant rated the shockwave as their favorite, while six rated it as their least favorite.}
    \Description{Horizontal stacked bar charts visualizing forced ranking results. Surveyor: 5 ranked it as \#1, 2 as \#2, and 2 as \#3. Menu: 4 as \#1, 4 as \#2, and 1 as \#3. Shockwave: 0 as \#1, 3 as \#2, and 6 as \#3.}
    \label{fig:forcedranking}
\end{figure}

Figure \ref{fig:forcedranking} shows the results of the forced ranking we asked participants for at the end of the study. Five and four participants ranked Surveyor and the audio menu, respectively, as their favorite tool of the three. No participant ranked the shockwave as their favorite tool. At the bottom of the rankings, six participants ranked the shockwave as their \textit{least} favorite tool. Two participants ranked Surveyor as their least favorite, and one participant ranked the audio menu as their least favorite. 

We report specific findings from our study in the following subsections. In Section \ref{sec:results:surveyor}, we discuss findings with respect to Surveyor and its ability to facilitate exploration and discovery. In Sections \ref{sec:results:surveyor-v-menu} and \ref{sec:results:surveyor-v-shockwave}, we compare Surveyor with the simple audio menu and the shockwave, respectively.

\subsection{Surveyor}
\label{sec:results:surveyor}

\subsubsection{Surveyor granted participants a strong sense of agency.} Participants generally felt that Surveyor granted them with a high sense of agency in exploring environments and discovering new items. Immediately after completing a level with Surveyor, P8 stated that they felt like they were on "top of the world." When asked to elaborate, they said:

\begin{quote}
    \textit{"[Surveyor] gives you a lot of confidence to be dropped at a random place and say, 'OK, I can just scan around me and see what's there, and I'll know if there's any objects around me and where else I can go to see new stuff.' [...] Like, I have a game plan in here."} --- \textbf{P8}
\end{quote}

Three participants explicitly noted the word "confidence" when talking about Surveyor, indicating how the tool can provide players with a "game plan" as they roam through the level. P9 explicitly noted that this was due to the multiple options he had access to for making sense of the level:

\begin{quote}
    \textit{"The thing that I liked about [Surveyor] is that I could feel what's around me [using the stick] and on top of that, the sub-menus were showing me what I hadn't explored and what I still need to explore. [...] So there were a lot of ways to see the level."} --- \textbf{P9}
\end{quote}

\begin{figure}
    \centering
    \includegraphics[width=0.45\textwidth]{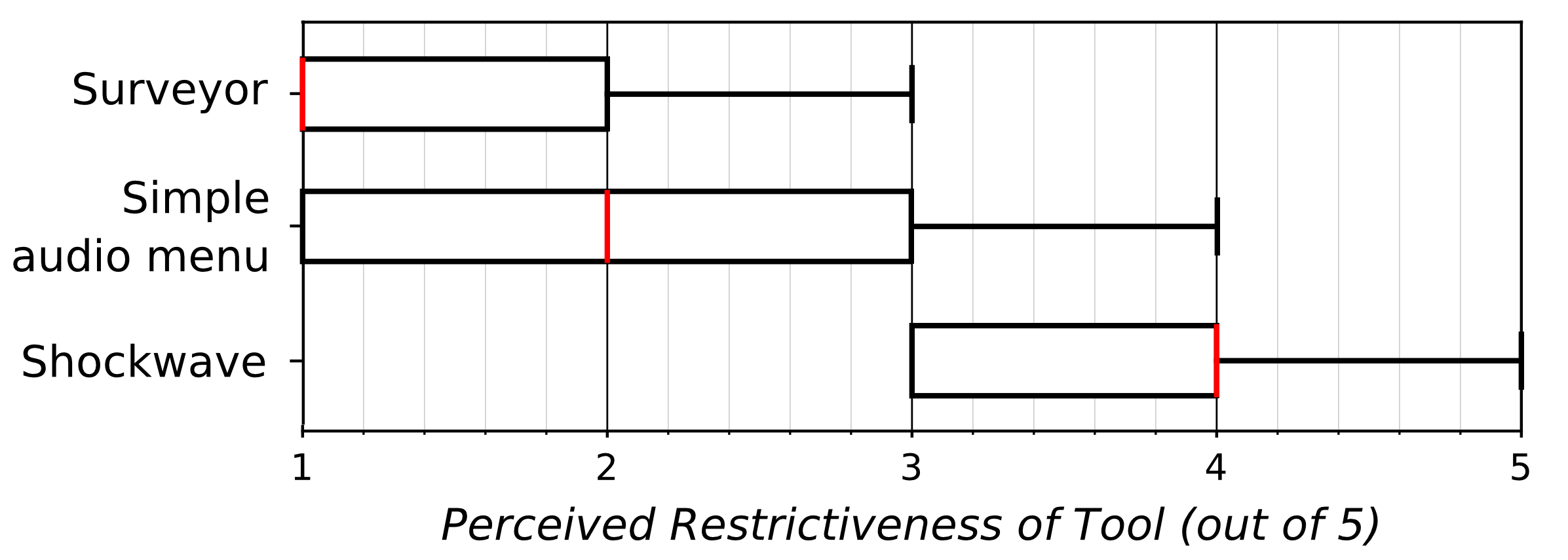}
    \caption{Distribution of scores indicating how restrictive participants perceived each tool to be. After using each tool in a level, participants were asked to rate its restrictiveness on a 5-point Likert scale, where 1 indicated that the tool felt "not at all restrictive" and 5 meant that it felt "extremely restrictive."  Surveyor received the lowest median restrictiveness scores at 1, followed closely behind by the simple audio menu at 2. The shockwave received the highest median scores with 4.}
    \Description{Box plots of perceived restrictiveness of each tool. Surveyor's max perceived restrictiveness as 3 out of 5. For the menu, it was 4 out of 5, and for the shockwave, it was 5 out of 5.}
    \label{fig:restrictivenessbox}
\end{figure}

Indeed, as Figure \ref{fig:restrictivenessbox} shows, Surveyor had the lowest median perceived restrictiveness out of all three tools with a median of 1 (meaning participants found Surveyor to be "not at all restrictive") versus 2 and 4 for the simple audio menu and the shockwave, respectively, on a 5-point Likert scale. Furthermore, six participants noted that Surveyor's tools provided them with complete freedom to explore around the level as they wish. 

\begin{figure*}
    \centering
    \includegraphics[width=0.95\textwidth]{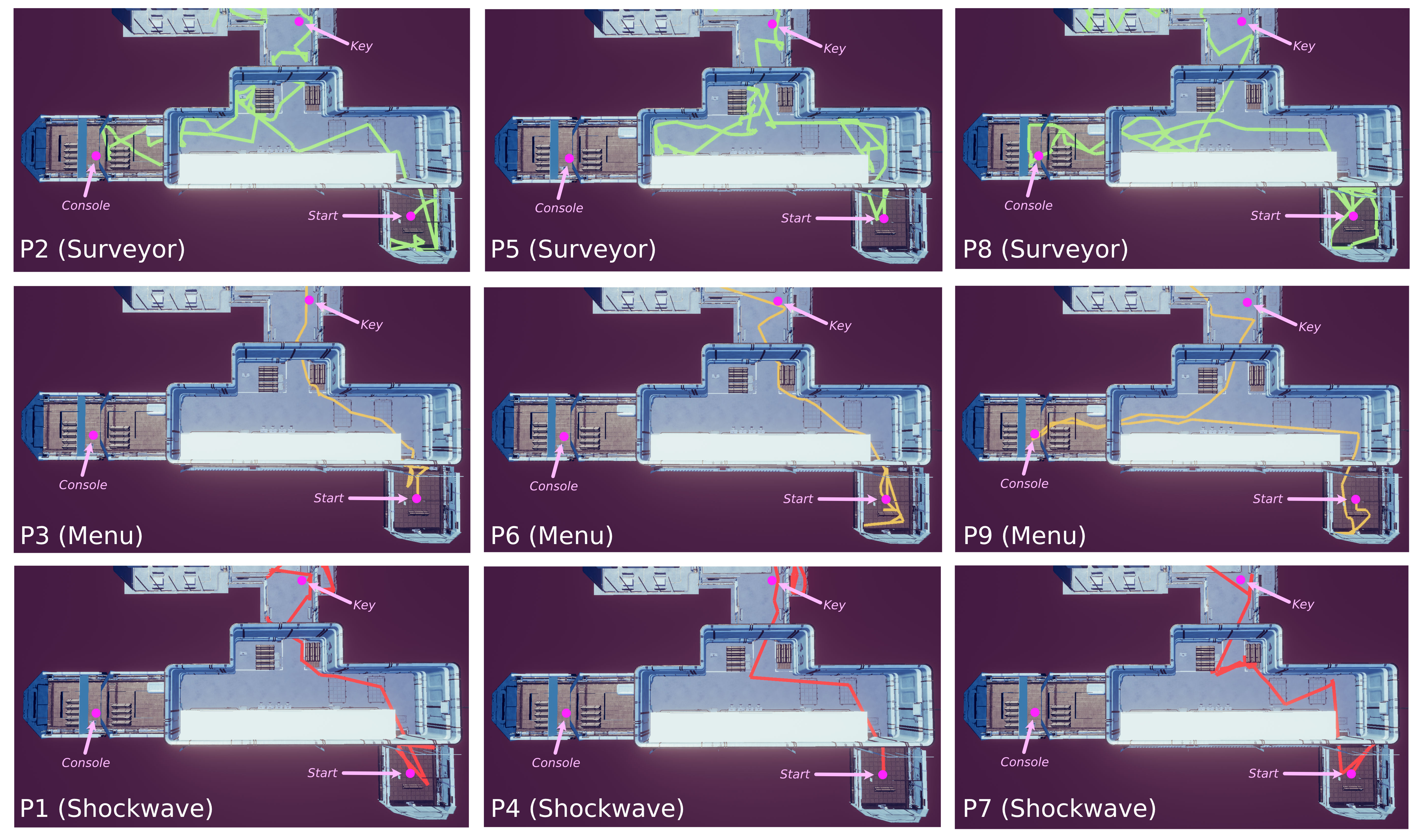}
    \caption{Paths of all nine participants at the start of Level 2. Players started in the room at the lower-right corner and had to collect the first key at the very top. The room at the far left is a "storage room" with a log entry console. Participants who used Surveyor tended to explore more of these rooms, with two participants finding the console and one reaching the door to the storage room before turning around. Only one participant who used the menu and none who used the shockwave entered the storage room. \textit{Path colors: Surveyor = green, simple audio menu = orange; shockwave = red.}}
    \Description{Nine copies of a top-down view of an area. The difference between the copies are the paths overlaid on top of the area.}
    \label{fig:paths}
\end{figure*}

These sentiments are corroborated when looking at the paths participants took to get through the level. Figure \ref{fig:paths} shows some of these paths. Note how participants who used Surveyor tended to explore more of the areas they passed through. Their paths were much less linear than the paths of those who used the simple audio menu or the shockwave, indicating that Surveyor tended to encourage players to explore the level more. 

P5 was particularly enthusiastic about this aspect of Surveyor. Although P5 was born with a non-functional eye, their other eye only gradually lost vision --- making them the only participant in our study to have had sighted experience with video games. In the post-study, P5 mentioned Surveyor as being closest to the sighted experience of the three tools he tried:

\begin{quote}
    \textit{"[Surveyor] is my favorite. Just being able to look around, know where you can go [...] The closest that you can get to the same experience as a sighted player is with Surveyor because you can look in a direction and see what's there and you can decide where you want to go from there. That's how you're supposed to do it."} --- \textbf{P5}
\end{quote}

\subsubsection{Some participants found Surveyor to be cumbersome to use.}

In general, participants felt that Surveyor elevated parts of the world that did not particularly matter to the objective at hand, and this was indeed by design: Surveyor highlights areas that the player has not yet explored, and it is very possible that some of these regions may not hold anything of interest to the user. For most, this was fun because it added an element of surprise to their exploration, and this closely tracks with what it means for a game to promote "discovery." However, for two participants, this approach felt cumbersome. P2 was one of them. They believed that this approach forced them to explore areas a little too much:

\begin{quote}
    \textit{"With Surveyor, it's like, you spin the stick around, but you're not getting everything --- you're only getting [what's around you], and it just forces you to feel out the rest of the room."} --- \textbf{P2}
\end{quote}

P6, the other participant who found Surveyor cumbersome, noted that the relative uniformity in how Surveyor presents unexplored patches forced them to randomly pick a patch and "hope for the best." Both participants' sentiments suggest that, although a tool like Surveyor provides a lot of choice in how players explore, some players may find this level of choice to be overwhelming. We discuss possible ways to improve these players' experience in Section~\ref{subsec:discussion--hints}.

%%%%%%

\subsection{Surveyor vs. the Simple Audio Menu}
\label{sec:results:surveyor-v-menu}

\subsubsection{The simple audio menu was easier to use than Surveyor.} Participants generally felt that the simple audio menu was easier to learn and use than Surveyor. Seven participants explicitly noted that Surveyor had a higher learning curve than the audio menu. Speaking about the audio menu, P5 noted its simplicity:

\begin{quote}
    \textit{"There is something to be said for just going straight down a list of 'here's all the things I could look at' [...] without having to worry about whether it's unexplored or not or which sub-menu I have to go into to find what I want to find. There's something to be said for just the simplicity of a single menu."} --- \textbf{P5}
\end{quote}

\begin{figure}
    \centering
    \includegraphics[width=0.45\textwidth]{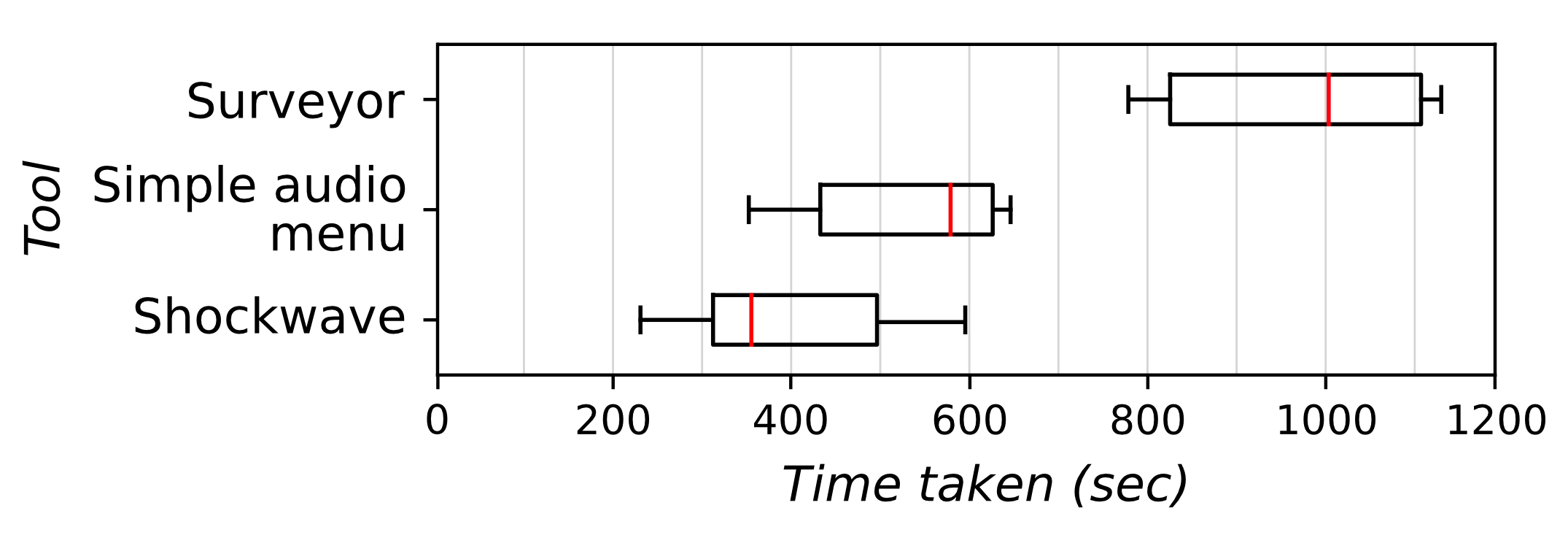}
    \caption{Distribution of level completion times with each tool. Levels where participants used Surveyor had, by far, the highest completion times (with a median of 1003 seconds) when compared to the simple audio menu (with a median of 585 seconds) and the shockwave (with a median of 355 seconds).}
    \Description{Box plots of level completion times with all 3 tools.  Surveyor had a minimum time of 782 seconds and a max of 1130 seconds. The menu had a minimum time of 363 seconds and a max of 642 seconds. The shockwave had a minimum of 234 seconds and a max of 590 seconds.}
    \label{fig:completiontimes}
\end{figure}

This is in contrast to Surveyor which included more moving parts (in terms of including the right stick \textit{and} the sub-menus) and encouraged participants to explore the levels more. This is very apparent in the time it took for participants to complete the levels. Figure \ref{fig:completiontimes} shows the distribution of times it took to complete the levels using each tool. It is clear that levels with Surveyor took \textit{much} longer (mean = 979 sec.) than levels with the audio menu (mean = 578 sec.) or the shockwave (mean = 234 sec.).

\begin{figure}
    \centering
    \includegraphics[width=0.41\textwidth]{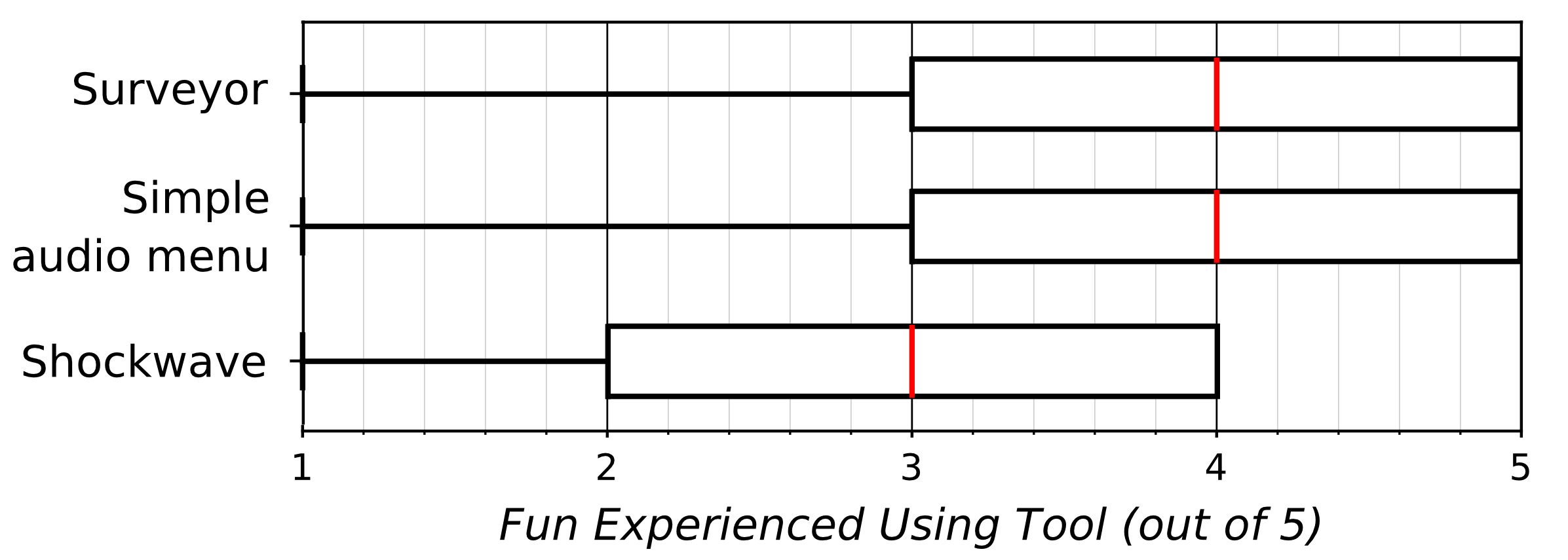}
    \caption{Distribution of scores indicating how fun playing the levels felt using each tool. After using each tool in a level, participants were asked to rate fun on a 5-point Likert scale, where 1 indicated that playing the level with this tool felt "not at all fun" and 5 meant that it felt "extremely fun."  Surveyor and the simple audio menu received identical distributions of scores from participants.}
    \Description{Box plots of perceived fun using all 3 tools. Surveyor and the simple menu had median scores of 4 out of 5. The shockwave had a median score of 3 out of 5.}
    \label{fig:funbox}
\end{figure}

Participants clearly spent more time in the levels using Surveyor, though this did not seem to negatively affect their perception of how fun the levels were when compared to the simple menu. Figure \ref{fig:funbox} shows that Surveyor and the simple audio-based menu received identical distributions of scores for perceived fun while using the tool --- both received a median of 4 ("very fun") on a 5-point Likert scale. Analyzing participants' sentiments on Surveyor and the audio menu yield two schools of thought on "fun" with both tools.

\subsubsection{Using the simple audio menu felt more "relaxing" but less fun.} As we mentioned in Section \ref{sec:results:surveyor}, with Surveyor, participants enjoyed being able to roam around the levels, exploring them, and discovering items, and this is despite them spending far more time within these levels using Surveyor than with the menu or the shockwave. However, we also saw that, for some, Surveyor forced players to explore areas even when they did not necessarily want, leading some of them to be annoyed by the approach.

With the simple audio menu, some participants appreciated just being able to immediately see what was in the room, and they found that ease to be fun it and of itself. P3 mentioned the following when talking about how fun the menu was right after trying it:

\begin{quote}
    \textit{"[The menu] was pretty easy and relaxing. I was able to finish the level pretty fast."} --- \textbf{P3}
\end{quote}

P3's sentiment points to this "relaxing" aspect as an important component for some participants of having fun while using these tools. However, for others, this ease and simplicity made the game much less interesting. P5 said the following after trying the menu --- in P5's case, our counterbalancing scheme had them try the menu \textit{after} trying Surveyor:

\begin{quote}
    \textit{"Obviously, I'm not timing myself, but I felt like I went through it faster [than with Surveyor]. But fast doesn't always mean better. The whole thing just felt very, very sterile and very soulless. It was just, 'Here's what you have access to, and if you want to go there, press a button,' and sure, you have to move, but that's it."} --- \textbf{P5}
\end{quote}

Four other participants mentioned similar sentiments. P4, for example, said that they could easily see what was in the area using the menu but that it took away the element of surprise "because you always know what you're going to." Participants' sentiments on the pros and cons of both tools yield insights into the role that a system like the simple audio menu can play within exploration assistance, and we discuss this further in Section~\ref{subsec:discussion--submission}.

%%%%%%

\subsection{Surveyor vs. the Shockwave}
\label{sec:results:surveyor-v-shockwave}

\subsubsection{Most participants disliked the lack of exploration granted by the shockwave.}

In stark contrast to both Surveyor and the simple audio menu, no participant placed the shockwave as their favorite. In fact, as the forced rankings in Figure \ref{fig:forcedranking} show, six participants rated the shockwave as their \textit{least} favorite. Participants' sentiments on the shockwave reveal their thoughts on mainstream game accessibility tools for BLV players --- especially those found in \textit{The Last of Us} games, after which our shockwave utility was modeled.

In terms of positives, all nine participants felt that, by using the shockwave, they were able to complete the level quickly, and this is confirmed by Figure \ref{fig:completiontimes}, which shows that the shockwave facilitated the lowest level completion times of all three tools we tested.

Despite this speed, however, participants did not feel they understood or felt immersed within the environment when using the shockwave. P1 stated the following:

\begin{quote}
    \textit{"There was just not enough information for me to feel like I was present in the world. I was just like trying to find stuff and hoping for the best. I didn't know what I was even really looking for except for a very general concept of the category of things. [...] And I didn't get a chance to look at any decorative objects. Seriously, I didn't know what I was looking at."} --- \textbf{P1}
\end{quote}

Like its counterpart within \textit{The Last of Us} games, our shockwave utility only emits pinging sounds when the shockwave hits objects in the environment. This is clearly not enough information for a BLV player to understand what the object is other than its general category by looking at the currently active filter.

This lack of detail alongside the shockwave's "golden path" implementation prevented participants from exploring the environment, and this is especially obvious in the number of "log entry consoles" that participants activated as they moved through the levels. Each level had three consoles that played clips narrating the backstory of the abandoned space station; these consoles were decorative objects and finding them was not necessary to complete the level.

With Surveyor and the simple audio menu, participants activated a median of three consoles (meaning that the median run with each tool had participants activate all available consoles). Participants could either look for consoles in unexplored patches (with Surveyor) or select the console from the menu (with the simple audio menu).

With the shockwave, however, participants activated a median of \textit{zero} consoles. It should be noted that three participants (P2, P7, and P8) each activated one console in the shockwave condition. However, all three of them noted that this was by complete accident --- while they followed a beacon, they deviated from the course and ended up in proximity of a console. P2 noted that the shockwave's pre-determined golden path prevented them from finding any consoles and learning more about the space station's backstory --- this made them lose interest in the game:

\begin{quote}
    \textit{"The only console I hit was by accident. I didn't even get the rest of the story, which was annoying me. But at that point I was like, 'I can't find consoles. The system doesn't allow me to do it. So why even bother?' [...] I have nothing good to say about this form of navigation. I hate this so much."} --- \textbf{P2}
\end{quote}

\subsubsection{Many participants disapproved of similar tools for BLV players within mainstream games.}

Five participants (including P2) expressed their disapproval of the tool within \textit{The Last of Us} games. Participants noted that, within these games, the "enhanced listen mode" forced participants to follow the golden path, and feedback from the mode's shockwave-like utility provided little information about objects outside of it. This granted players with very little opportunity to explore the game's richly detailed environments --- environments that are fully accessible to sighted players. P2, in particular, said that they did not finish either of \textit{The Last of Us} games because the tools that the games provided to BLV players made them feel "like an automaton" since they could only follow the game's prescribed path and not do anything else.

In contrast, some participants felt that Surveyor offered them the ability to do anything they wanted, including the ability to look for other items within the world as they wished.  P5, in particular, noted that Surveyor allowed them to make mistakes as opposed to the shockwave, which led them along a more "perfect" path:

\begin{quote}
    \textit{“I liked being able to make decisions, you know --- even if that leads me to making a wrong decision because I'm human, you know. [...] That's what gaming is about: It's about the decisions that you make having an impact on what happens. And I liked that [Surveyor] didn't stand in the way of that happening.”} --- \textbf{P5}
\end{quote}

P5's sentiment implies that experiencing the effects of one's actions is a crucial aspect of facilitating a player's agency with a game, and indeed, prior work has confirmed this~\cite{Cole2021}. Exploration assistance tools, like Surveyor, can facilitate these experiences, thus fostering a greater sense of agency and fun in players.

%% file: sec08-discussion.tex
Our study findings reveal several implications for future exploration assistance systems, especially for those that seek to promote a sense of exploration and discovery for BLV players within video games. In this section, we reflect on these implications and propose avenues for future work.

%%%%%%

\subsection{Facilitating greater agency and joy through in-game exploration assistance.}

Our findings confirm that providing in-game navigation assistance tools that \textit{also} foster exploration and discovery can grant BLV players a greater sense of agency over what actions they take within a game \textit{and} a sense of joy as a result of those actions. Participants' sentiments on the simple menu and the shockwave show that current navigation assistance tools fail to facilitate both: With the shockwave, participants felt they had no control over what they did within the game. The menu afforded some control to players in terms of what places they could move toward, but many participants felt that the game world lacked an element of surprise.

With Surveyor, however, we observed how exploration assistance \textit{did} facilitate these feelings in many of our participants. By allowing them to scan the area as they wish, move through the environment freely, and rapidly query their current exploration status, players were able to understand where they were, make decisions according to that understanding, and experience the outcomes of those decisions --- a process that plays a key role in granting autonomy to sighted players in many games \cite{Tyack2021, Ryan2006}. Future assistive tools within games with detailed environments should strive toward facilitating exploration assistance to improve BLV players' experience within those games.

%%%%%%

\subsection{Incorporating "hints" into exploration assistance.}
\label{subsec:discussion--hints}

We designed Surveyor to operate on a zero-initial-knowledge principle, where neither the tool nor the player possesses information about what lies within unexplored areas until they are "seen." However, some participants --- namely, P2 and P6 --- felt compelled to explore entire areas using Surveyor, even when they wanted to stop and move on, decreasing their overall satisfaction with the tool. This issue stemmed, in part, from the fact that all unexplored patches are presented uniformly, leaving players unaware of what an unexplored area contains or \textit{if} it holds anything of interest.

Sighted exploration works on a similar principle where the user knows they have not explored certain areas and makes their way to these areas to see what lies within them. Yet, within games, sighted players not only use vision to gain information about their environments, they also use supplementary tools, such as compasses and minimaps. Information from these tools, which can include unexplored area size and the presence of nearby cues, can serve to nudge players in a given direction~\cite{Thorn2018, Bork2019, Zagata2021}. These information cues can be viewed as contextual "hints" that may streamline the exploration process, though this streamlining is not guaranteed.

Providing similar hints may also help BLV users in their exploration process and further enhance their sense of agency. Indeed, prior work has found that offering hints was valuable in helping BLV users to better structure their tactile explorations of digital images using touchscreens while preserving their agency~\cite{Nair2023b}. Although this prior work focused on images, future work looking at in-game exploration assistance tools could explore how a similar hints-based approach might manifest and assist BLV players in making more informed decisions about how to approach exploration --- using findings from prior work looking at what supplementary information BLV users find important while exploring~\cite{Nair2022a, Jain2023}. Such systems, however, should take care to preserve a sense of discovery and not spoil the locations of items and places.

%%%%%%

%%% DONE %%%
\subsection{Granting players the option to "reveal" environments for an easier experience.}
\label{subsec:discussion--submission}

Prior work looking at BLV players' gaming experiences has decried menu- and grid-based systems as over-simplifying games and making them less fun for BLV players when compared to sighted players~\cite{Andrade2019, Nair2021, Smith2018}. Many of our participants' sentiments on the simple audio menu corroborate this view. Our findings, however, \textit{also} indicate that menu-based systems may have a place in facilitating less intense experiences for players should they desire them. Indeed, many participants acknowledged that the ease-of-use of the simple audio menu made it an attractive option for quickly seeing what was in an environment and moving from point to point. Participants' desire for an "easier" experience points to allowing this ease-of-use as an important component of facilitating fun when exploring. In fact, Gon\c{c}alves et al. found that some BLV players prefer to enjoy games in a more passive manner, perhaps to enjoy a game's sound design~\cite{Goncalves2023b}.

A question, then, arises as to how we can tweak in-game exploration assistance tools so that they not only give players the ability to discover items on their own but also give players the option to skip the exploration and immediately get a list of all objects and their locations within the environment. In Surveyor's case, one way is to give the player the option to \textit{automatically} explore the rest of the area on the press of a button. This would fill out the seen landmarks sub-menu, effectively reducing Surveyor's user interface to a simple audio menu that lists all objects. Such functionality would give players the option to actively explore the area by learning about unexplored regions \textit{or} passively learn about it using a complete list of landmarks. Future in-game exploration assistance tools could explore granting similar options that players could use if they so desire.

%%%%%%

\subsection{Future work: Surveyor and exploration assistance in other contexts.}

The concepts we introduce within this work mainly apply to games where \textit{discovery} is important. Recall from Section \ref{subsec:gdtheory}  that discovery is only one of several categories game designers use to formalize what makes a game "fun," and different types of games prioritize these categories to varying extents~\cite{RiotGames2019, Garneau2001, Hunicke2004}. Among games that primarily facilitate discovery, Surveyor, in its current form, works for scanning environments in a top-down manner, rendering it well-suited for 3D environments or top-down 2D settings. Yet, not all discovery-focused games are like this --- an example is \textit{Super Mario Bros.}, a 2D platforming game that encourages players to look for coins and special items~\cite{NintendoR&D41985}. Future work can look at adapting our principles for in-game exploration assistance for these non-3D, non-top-down game contexts.

Another area of future work may involve investigating Surveyor-like exploration assistance tools within the physical world. Although not targeted toward BLV users, an analogous system is the now-defunct Zenly, a social location-sharing app that mapped users' visited places within the world and "uncovered" them on a map, encouraging them to explore places they had not yet uncovered~\cite{Zenly2022}. Existing navigation assistance systems for BLV users, however, are typically limited to providing turn-by-turn directions and may offer only basic information about what lies in the person's immediate environment~\cite{Sato2017, MicrosoftResearch2018}. Jain at al. noted a need for more exploration-oriented assistance systems for BLV users~\cite{Jain2023}. Thus, future work may investigate how a Surveyor-like exploration assistance system can be adapted to the physical world. For example, adding Surveyor's precise tracking and guidance mechanisms may encourage BLV users to explore not only expansive urban areas but also more compact environments, such as university campuses, museums, and parks --- fostering more enjoyable exploration experiences within physical environments.

%% file: sec09-limitations.tex
We acknowledge several limitations with our work. As with many studies that include people with disabilities --- especially those who are BLV --- we had a relatively low number of participants. The sentiments of our nine participants may not necessarily be representative of the BLV gaming community at large. It is important to note that Surveyor is \textit{one} possible manifestation of the design goals we found through our formative interviews and that there may be other goals we did not consider. More work is also needed to determine how we can better adapt tools for BLV users with other disabilities: For example, although we had two participants with hearing impairments, their hearing issues were not an impediment to them using the tools in our main study. However, there may be tweaks we can make for users with more severe hearing impairments or for users with physical impediments that may prevent effective use of a standard game controller.

%% file: sec10-conclusion.tex
In this work, we designed, built, and evaluated Surveyor, an exploration assistance tool for virtual environments that grants BLV users a level of discovery greater than that offered by existing navigation assistance tools. Through formative interviews, we identified specific design objectives that Surveyor needed to fulfill in order to enable BLV players to effectively explore game worlds. We then designed and implemented Surveyor and evaluated its effectiveness in promoting exploration and discovery, comparing it to approaches found within existing accessible games through a study involving nine participants.

During the formative interviews, we found that BLV players desired to discover items while physically moving through virtual worlds as opposed to simply querying lists and grids. They also sought ways to keep track of explored and unexplored areas, use audio beacons for efficient navigation to specific locations, and have a tool that offered multiple approaches to understanding and navigating through the environment. To address these design objectives, we built Surveyor and conducted an evaluation against two established in-game accessibility approaches: a simple audio menu (representing popular approaches for representing worlds within games designed for BLV players) and a shockwave utility (representing mainstream video game BLV accessibility features). Our findings revealed that Surveyor granted players a high degree of agency in exploring environments and discovering new items despite its initial learning curve; the simple audio menu allowed players to finish levels quickly despite diminished surprise during item searches; and the shockwave failed to provide players with a sense of enjoyment or immersion within the environment.

We hope that by understanding what BLV gamers prefer when exploring game environments and by creating tools to make their gaming experience more enjoyable, we can inspire game designers and developers to include more accessibility features in their games --- allowing BLV players to have more fun with these games and enjoy gaming experiences similar to those of sighted players.